\documentclass[useAMS]{mn2e}
\usepackage{graphicx}

\title[]{}
\author[]{}

\title[Dynamical friction in dwarf galaxies] {Theoretical lower limits on sizes of ultra faint dwarf galaxies
from dynamical friction} 
\author[X. Hernandez] {X. Hernandez \\Instituto de Astronom\'{\i}a,
  Universidad Nacional Aut\'{o}noma de M\'{e}xico, Apartado Postal
  70--264 C.P. 04510 Ciudad de M\'exico, M\'exico.} 

\date{Released 17 May 2016}

\pagerange{\pageref{firstpage}--\pageref{lastpage}} \pubyear{2015}

\begin{document}

\label{firstpage}

\maketitle

\begin{abstract}
Dwarf spheroidal galaxies are the smallest known stellar systems where under Newtonian interpretations,
a significant amount of dark matter is required to explain observed kinematics. In fact, they are in this sense
the most heavily dark matter dominated objects known. That, plus the increasingly small sizes of the newly
discovered ultra faint dwarfs, puts these systems in the regime where dynamical friction on individual
stars starts to become relevant. We calculate the dynamical friction timescales for pressure supported
isotropic spherical dark matter dominated stellar systems, yielding $\tau_{DF} =0.93 (r_{h}/10 pc)^{2} 
(\sigma/ kms^{-1}) Gyr$, { where $r_{h}$ is the half-light radius}. For a stellar velocity dispersion
value of $3 km/s$, as typical for the smallest of the
recently detected ultra faint dwarf spheroidals, dynamical friction timescales becomes smaller than the
$10 Gyr$ typical of the stellar ages for these systems, for $r_{h}<19 pc$. Thus, this
becomes a theoretical lower limit below which dark matter dominated stellar systems become unstable to
dynamical friction. We present a comparison with structural parameters of the smallest ultra faint dwarf 
spheroidals known, showing that these are already close to the stability limit derived, any future 
detection of yet smaller such systems would be inconsistent with a particle dark matter hypothesis.
\end{abstract}


\section{Introduction} \label{intro}

Over the past few years the number of ultra faint dwarf spheroidal galaxies (dSphs) has grown significantly, 
turning them from a handful of peculiar objects, to a class of increasingly well studied systems, e.g. 
Belokurov et al. (2007), Martin et al. (2008). Recent studies have gone from their photometric identification 
to detailed dynamical and stellar population characterisations of this small systems, e.g. Kirby et al. (2015a), 
Voggel et al. (2016). Their most salient feature is their extremely large mass to light ratios, frequently in 
the hundreds and sometimes of upwards of a thousand, when interpreting these systems as equilibrium structures 
under Newtonian gravity, e.g. Gilmore et al. (2007), Kirby et al. (2015b). Regarding their stellar populations, 
all are consistent with low metallicities and very old stellar ages of order $10 Gyr$, e.g. Hernandez et al. 
(2000), Drlica-Wagner et al. (2015). This last and the relaxation timescales of their stellar populations of 
many Hubble times, strongly suggest interpreting ultra faint dSphs as equilibrium systems.

{  Under the current $\Lambda CDM$ structure formation scenario, the most salient and persistent missmatch
with observations occurs at the smallest scales, where models appear to over-predict the number of satellites
a large galaxy like the Milky Way should have, e.g. Moore et al. (1999), Kang et al. (2016). Also, it is at the
smallest scales of the dwarf spheroidal systems where tests of the dark matter hypothesis become cleaner and more
sensitive, given the absence of gas and the presence of well studied stellar populations, they have been recently
used to attempt predicting direct dark matter annihilation signals e.g. Bonnivard et al. (2015) and constraining
modified gravity theories e.g. Hernandez et al. (2010). Over the past few years, use of large surveys have
significantly improved the observational picture, with the discovery of many new dSphs, later selected
as targets for detailed spectroscopic studies e.g. the Sloan Digital Sky Survey, the Panoramic Survey Telescope
and the now public Dark Energy Survey e.g. Belokurov et al. (2007), Koposov et al. (2015) and Martin et al. (2016).
}

As the existence of ultra faint dSphs had not been predicted under standard CDM structure formation scenarios, 
their detection and study remain a largely empirical endeavour, where no limits on their structural parameters 
are expected {\it a priori}. One of the most interesting developments has been the discovery of increasingly smaller 
systems, with half-light radii which have decreased from the hundreds of pc of the first ultra faint dSphs 
(e.g. Martin et al. 2008), to several tens for the latter ones (e.g. Drlica-Wagner et al. 2015), to just $19 pc$  
for the recently discovered Draco II, with a velocity dispersion of $2.9 \pm 2.1 km/s$, and a resulting mass to 
light ratio of between and 80 and 1,600, Laevens et al. (2015), Martin et al. (2016). {  As the sizes of detected
dSphs have gone down, their inferred numbers of stars have fallen from the millions of the classic dwarfs
to just thousands or even a few hundreds in the recent ultra faint dSphs, e.g. Misgeld \& Hilker (2011).}
In this paper we derive a theoretical limit on the smallest size of a stable dark matter dominated spherical
stellar system, through dynamical friction considerations.

Dynamical friction of a dark halo on individual stars is generally ignored, due to the extremely large decay 
timescales that result, of many orders of magnitude larger than the Hubble time, for typical galactic parameters 
of velocities $v \sim 100 km/s$ and sizes of $R \sim 5 kpc$. However, dynamical friction decay timescales, 
$\tau_{DF}$, scale with $\tau_{DF}\propto R^{2}v$. Thus, for systems with sufficiently small half-light radii 
and internal velocities, $\tau_{DF}$ will eventually become smaller than the typical ages of stellar populations 
in ultra faint dSphs. In this paper we derive dynamical friction timescales for spherical pressure supported 
isotropic and dark matter dominated systems, as all dSphs are inferred to be under Newtonian interpretations. 
A dynamical friction timescale results of $\tau_{DF}=0.93 (r_{h}/10 pc)^{2} (\sigma /km s^{-1}) Gyr$, 
where $r_{h}$ and $\sigma$ are the stellar half light radius and velocity dispersion of the stars in a dark 
matter dominated dSph, respectively. For typical $\sigma \sim 3 km/s$ values, as empirically determined for 
the smallest ultra faint dSphs known today, $\tau_{DF}$ becomes shorter than $10 Gyr$ at half light radii of 
$19 pc$. Systems smaller than this will therefore be unstable towards dynamical friction decay, within their 
assumed dark matter halos. The discovery of systems at around this value and smaller would therefore be inconsistent 
with the dark matter particle hypothesis, while being natural under MONDian gravity expectations, where no dark 
matter halo is envisioned, and hence no dynamical friction is to be expected for these objects, which incidentally, 
present $\sigma$ values in accordance with their observed stellar masses and simple MONDian expectations, e.g. 
Hernandez et al. (2010), Lughausen et al. (2014).

Dynamical friction consistency arguments have been used before to provide
constraints on dark matter halo properties. For example, the first study
suggesting constant density cored dark matter halos in dSphs was Hernandez
\& Gilmore (1998), through requiring dynamical friction timescales for
globular clusters in the Sagittarius dwarf to be longer than the age of
the system, or more recently, Sanchez-Salcedo et al. (2006) applying the
same argument to a number of classical dSphs.

In section 2 we derive the dynamical friction timescale for spherical pressure supported dark matter dominated 
systems, through tracing, not the decay of the angular momentum, as customary for circular orbit dynamical friction 
decay calculations, but the decay of the potential energy of the stellar population characterised by its $\sigma$ 
value. In section 3 we present a comparison of this $\tau_{DF}$ timescales to the typical ages of the stellar 
populations in observed dSphs, for structural and kinematic values of some of the smallest ultra faint dSphs 
known to date. Section 4 presents a short discussion and concluding remarks.

\section{Dynamical friction solution}

A star moving through a distribution of dark matter particles will gravitationally deflect each particle it
encounters, with a net effect resulting in a dynamical frictional drag force, $F_{DF}$, acting upon the star in the 
direction opposite its motion. Considering a body of mass $M$ moving at velocity $V$ through a distribution of dark 
matter particles with density $\rho$ and Maxwellian velocity dispersion $\sigma$, starting from the Chandrasekhar formula 
(Chandrasekhar 1943), we obtain the following classical expression for the frictional 
drag on $M$:

\begin{equation}
F_{DF}=-4 \pi ln\Lambda G^{2} \rho M^{2} \frac{A(X)}{V^{2}}. 
\end{equation}

In the above it is assumed that the mass of the dark matter particles is much smaller than that of the star.
$ln\Lambda$ is the Coulomb logarithm of the problem, $\Lambda=b_{max} V^{2} / (G M)$, where $b_{max}$
is a maximum impact parameter relevant to the problem, generally taken as the size of the dark matter particle
halo in question. Also, $A(X)=erf(X)-2Xe^{-X^{2}}/\pi^{1/2}$, where the variable $X=V/(\sqrt 2 \sigma)$.
For stars moving at the circular equilibrium velocity of an isothermal dark matter halo, $X=1$, $A(X)=0.428$ and
$\rho=V^{2}/(4 \pi G R^{2})$, $R$ the orbital radius of the star in question, yielding:

\begin{equation}
F_{DF}=-0.428 ln\Lambda \frac{G M^{2}}{R^2}.   
\end{equation}

\noindent The preceding development is well known, e.g. Binney \& Tremaine (1987). 
The standard way of proceeding now is to consider the loss of angular momentum of the particle in question due 
to $F_{DF}$, assume circular orbits, and obtain an equation for the reduction of the orbital radius with time.
As we wish to explore the problem for stars in dSphs, systems where stellar orbits are not circular but rather dominated
by velocity dispersion, we introduce here an analogous approach based on tracing the potential energy loss of 
the star undergoing dynamical friction.

The loss of potential energy for the star in moving through a distance $dx$ is now $dw=F_{DF} dx$. Dividing
by $dt$ we obtain,

\begin{equation}
\frac{dw}{dt}=F_{DF} \frac{dx}{dt} = V F_{DF}. 
\end{equation}

Assuming still circular orbits in an isothermal halo characterised by a logarithmic potential, the loss of
potential energy for the star when its orbital radius changes from $R+dR$ to $R$ will be given by
$dw=M V^{2} ln(1+dR/R)$, which for tightly wound orbits and $dR<<R$ reduces to $dw=M V^{2} dR/R$
and hence,

\begin{equation}
\frac{dw}{dR}=\frac{M V^{2}}{R}.
\end{equation}

\noindent Since $dR/dt=(dR/dw)(dw/dt)$, we can now write the evolution equation for the orbital radius 
taking $dw/dt$ from equation (3) with $F_{DF}$ from equation(2) and $dR/dw$ from equation (4) above to
yield:

\begin{equation}
\frac{dR}{dt}=-0.428 ln \Lambda \frac{G M }{R V}.
\end{equation}

The above is exactly the same evolution equation which results from starting from equation(2) 
(Binney \& Tremaine 1987 eq. 7-25) and tracing the loss of angular momentum for the star assuming slowly inspiraling 
circular equilibrium orbits, and validates the approach introduced of tracing rather, the evolution of the potential 
energy of the star being followed. 

If we now return to stars supported by velocity dispersion in dSphs, changing $V$ for $\sigma_{*}$, and assuming
initially that the velocity dispersion of the stars equals that of the dark matter particles, $X=1/ \sqrt 2$, 
and $A(X)=0.2$. Going back to equation (1), the dynamical frictional force on the sample star is now:

\begin{equation}
F_{DF}=-2.5 ln\Lambda \rho \left( \frac{G M}{\sigma_{*}} \right)^{2}.
\end{equation}

\noindent As previously, $dw/dt=\sigma_{*} F_{DF}$, and hence the total rate of loss of potential energy for all the
$N$ stars in the dSph galaxy will be:

\begin{equation}
\frac{dW}{dt} =-2.5 \frac {ln\Lambda N \rho (G M)^{2}}{\sigma_{*}}  
\end{equation}

For a star of mass M at a radial distance $R$ within the constant density core of a dark matter halo of density $\rho$,
as inferred in general for dSphs (e.g. Goerdt et al. 2006), in moving a radial distance $dR$ , the change in potential 
energy is given by:

\begin{equation}
\frac{dW}{dR}=-\frac{4 \pi}{3} G M \rho R.
\end{equation}

\noindent Evaluating the above at the half-light radius, $r_{h}$, and using
 $dW/dt=(dW/dr_{h})(dr_{h}/dt)$ allows us to write for the full N stars:

\begin{equation}
\frac{dW}{dt}= 4.2 G M N \rho r_{h} \frac{d r_{h}}{dt}, 
\end{equation}

\noindent from which we can solve for the decay rate of the half light radius, taking $dW/dt$ from
equation(7) to get,

\begin{equation}
\frac{dr_{h}}{dt} =-0.6 \frac{ln\Lambda G M}{\sigma_{*} r_{h}}
\end{equation}

\noindent It is reassuring of the development presented that the above equation agrees with the classical expression 
for the decay rate of the orbital radius of a particle inspiraling within a dark matter halo along quasi-circular orbits,
exactly in all the physical dependencies, and to within a factor of order unity in the numerical coefficient, a minor
difference due to the slightly distinct physical conditions of the pressure supported system being treated here.

Integrating the above equation and defining $\tau_{DF}$ as the time taken for the stellar half-light radius to 
be reduced by a factor of two gives:

\begin{equation}
\tau_{DF}=0.63 \frac{\sigma_{*} r_{h}^{2}}{ln\Lambda G M},
\end{equation}

\noindent which in astronomical units for $1 M_{\odot}$ stars, yields the {  main} result of this section:

\begin{equation}
\tau_{DF}=\frac{0.14}{ln \Lambda}(\sigma_{*}/km s^{-1})(r_{h}/pc)^{2} Gyr.
\end{equation}

\noindent Again, the same physical scalings and order of magnitude than what results from following the loss
angular momentum for a particle in a circular orbit within an isothermal halo.
{  For a value of $ln \Lambda=15$, corresponding to typical values for ultra faint dSphs, the above equation
yields:$ \tau_{DF} =0.93 (r_{h}/10 pc)^{2} (\sigma/ kms^{-1}) Gyr$.}

Once a dominant dark matter halo has been assumed, $\tau_{DF}$ depends only on
$\sigma_{*}$ and $r_{h}$, and not explicitly on the dark matter density or mass,
as happens also for the classical inspiraling of circular orbits due to
dynamical friction within an isothermal halo, since both the dynamical friction attrition on the potential
energy of the target star, and the star's potential energy scale linearly with the dark matter density.
This is fortunate, as actual
M/L ratios (and hence total dark matter masses) for this smallest of systems
are subject to significant statistical fluctuations due to an incomplete
sampling of the initial mass function. When only a few hundred stars are
available, turning colours and luminosities into total stellar masses
through population synthesis results which assume a densely sampled IMF,
leads to systematic errors of factors of a few, as shown in Hernandez (2012).
Also for the reasons noted above, equation (12) will be fairly robust to the details of
the dark matter density profile.

Notice that from equation (10), the decay rate increases as the typical radius decreases, in fact, the time for 
$r_{h} \rightarrow 0$ through equation (10) is only 1.33 times the value given in equation (12). After $\tau_{DF}$
has elapsed, the stellar population rapidly evolves into a tight star cluster. {  This can be seen more clearly
by considering how the dark matter mass within the region occupied by the stars decreases as the typical
stellar radius goes down. For example, for evolution within a constant density dark matter halo, after
three $\tau_{DF}$, the dark matter mass within the typical stellar radius will have decreased by a factor
of $8^3=512$, bringing the mass to light ratio to within stellar values, i.e., the structure will 
no longer be dark matter dominated, and will appear more like a globular cluster than a dSph. Once the
potential energy is dominated by the stars, eq.(7) remains valid, but the stellar potential energy will be given by:

\begin{equation}
W=- 0.4 \frac{G(MN)^{2}}{r_{h}},
\end{equation}

\noindent e.g. Binney \& Tremaine eq. (4-80b). Differentiation w.r.t. $r_{h}$ of the above, and proceeding as in the
previous dark matter dominated case, yields:

\begin{equation}
\frac{dr_{h}}{dt}=-6.25 \frac{ln\Lambda G \rho r_{h}^2}{N\sigma_{*}} 
\end{equation}  

\noindent as the corresponding equation to eq.(10). Similarly, integrating and defining again $\tau_{DF*}$ as the time
required for $r_{h}$ to go down by a factor of two, leads to:

\begin{equation}
\tau_{DF*}=\frac{0.16 N \sigma_{*}}{ln\Lambda G \rho r_{h}},
\end{equation}

\noindent for the equation corresponding to eq.(11), which in astrophysical units reads:

\begin{equation}
\tau_{DF*}=\frac{40 N_{3}}{ln \Lambda} (\sigma_{*}/kms^{-1})(\rho /M_{\odot}pc^{-3})^{-1}(r_{h}/pc)^{-1} Gyr,
\end{equation}

\noindent where $N_{3}$ is the total number of stars in units of thousands. Again, for $ln\Lambda =15$
we obtain, $\tau_{DF*}=0.27 N_{3}(r_{h}/10 pc)^{-1}(\rho/M_{\odot}pc^{-3})^{-1}(\sigma_{*}/kms^{=1})$. By comparing
to the corresponding expression for the dark matter dominated phase, we see a similar scale for $N_{3}=1$
and $\rho = 1 M_{\odot}pc^{-3}$, parameters typical for ultra faint dSphs. Of course, the change in the potential
energy from dark matter dominated to a self gravitating population of stars, changes the radial dependence
on the evolution timescales to $R^{-1}$, such that the process is now self-limiting and the evolution timescales
now grow as the radius is reduced. As $r_{h}$ goes down, the process stops and a tight stellar cluster,
rather than a dark matter dominated dSph, would result. The transition between the two regimes will not be
abrupt, but such details lie beyond the scope of this first study, the intention of which is mainly to
raise awareness to the relevance of a physical process which has hitherto not been considered.

When the potential of the stars themselves begins to dominate over that of the dark matter, the hypothesis
leading to equation (10) breaks down, and the evolution reaches a phase described by the above equations.
Ultra faint dSphs with $\tau_{DF}$ shorter than their ages will be unstable structures which should appear
observationally extremely rare, under a particle dark matter hypothesis. 
}

\section{Comparison with observed Ultra Faint dSphs}

We begin by noting that for the classical dSphs, with structural parameters in the ranges:
$r_{h} \approx 500 pc$, $\sigma_{*} \approx 10 km/s$ and total dark halo sizes of order $10 kpc$
(e.g. Salucci et al. 2012), we obtain $ln \Lambda \approx 20$, and from equation (12), values of
$\tau_{DF}$ of about $10^{3}$ times the age of the universe. For this reason, dynamical friction on individual stars 
is generally ignored, even for such high inferred mass to light ratio systems as the classical dSphs, the effect is
generally entirely negligible.

However, the strong $r_{h}^{2}$ dependence seen in equation (12) suggests that in going to the newly discovered ultra
faint dSphs, interesting effects might appear. Typical values for the ultra faint dSphs are of order:
$r_{h} \approx 30 pc$, $\sigma_{*} \approx 5 km/s$, e.g. Martin et al. (2008). Assuming total dark halo sizes of 
$\approx 1 kpc$, results in $ln \Lambda =15$. Notice that the precise values for the total halo extent enter only 
into the estimate of $ln \Lambda$, and linearly in the logarithm, making all that follows highly robust to even order 
of magnitude uncertainties in the assumed total halo extent. Typical parameters for the more recently discovered ultra 
faint dSphs now yield $\tau_{DF} \approx 30 Gyr$. Again, if only half a dynamical friction timescale has elapsed over 
the lifetimes of the systems involved, we can still conclude the effect to be marginal. 

\begin{figure}
\includegraphics[width=9.0cm,height=7.0cm]{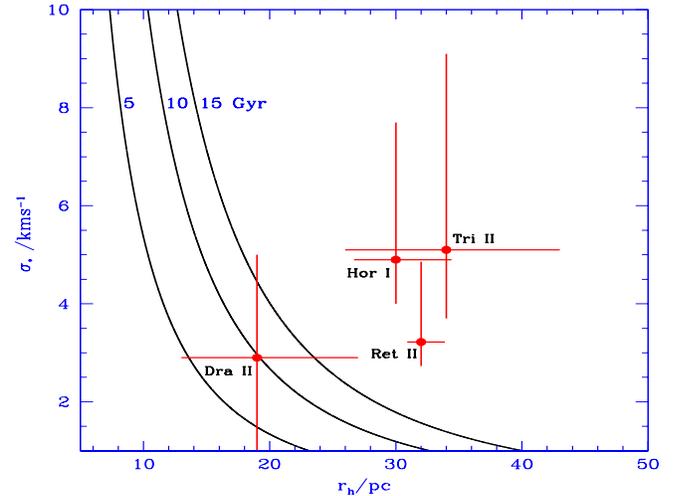}
\caption{Observed values of $r_{h}$ and $\sigma_{*}$ for various recently detected ultra faint dSphs, with
reported confidence intervals, see table for sources. Assuming $ln \Lambda=15$, the solid curves give the 
loci of $\tau_{DF}=5, 10$ and $15 Gyr$, left to right respectively. The smallest ultra faint dSphs appear 
already at the instability limit for dynamical friction shrivelling under a particle dark matter hypothesis.
A reduction of an order of magnitude in the assumed total dark halo radii would result in an increase in
$\tau_{DF}$ of $18\%$, the curves would correspond to $\tau_{DF}=5.9, 11.8$ and $17.7 Gyr$}
\end{figure}

The recently discovered Tuc V and Cet II reported by Drlica-Wagner et al. (2015) are even smaller, with half-light 
radii of 17 pc in both cases. Although no kinematic studies have been published for either, the position of Tuc V
on the magnitude-size plane in the region occupied by other well studied dSphs, together with the ellipticity observed, 
leads the authors to strongly identify the object as a galaxy. Its small stellar mass of only $\approx 500 M_{\odot}$
and known scalings for the ultra faint dSphs, suggest an internal velocity dispersion $<3km s^{-1}$, which would
result in $\tau_{DF}=8 Gyr$, for the central value of its inferred $r_{h}$. Given the inferred age of $10.9 Gyr$ for
Tuc V (Drlica-Wagner et al. 2015), this system is probably slightly beyond the instability limit for dynamical 
friction shrivelling. The case of Cet II is more ambiguous, and could turn out to be an outer globular cluster.

We now go to the most extreme confirmed case, the Draco II ultra faint dSph recently reported in Laevens et al. (2015) and
Martin et al. (2016). For this system a half light radius of only $19 pc$ was reported, with an observed velocity 
dispersion of $3 km/s$. Keeping an $ln \Lambda=15$ value, the resulting dynamical friction timescales for Draco II 
is of $\tau_{DF}= 10.2 Gyr$. We see that for Draco II, about one dynamical friction timescale has elapsed over the 
typical age of a dSph stellar population of $\approx 10 Gyr$, suggesting significant dynamical evolution should be 
expected, something which is not evident from the appearance of the system in question, which shows no distinct 
features when compared to the other ultra faint dSphs.

Figure 1 presents observed $r_{h}$ and $\sigma_{*}$ values for a collection of recently observed ultra faint dSphs, 
with confidence intervals in reported quantities shown.
Assuming a constant $ln\Lambda =15$ value {  we show} in the figure curves
of constant $\tau_{DF}=5, 10$ and $15 Gyr$, from left to right.
We see that Dra II is already close to the stability limit for dynamical
friction shrivelling, significantly over the limit if future studies were
to revise its size towards the lower ranges of its current reported
confidence interval. The same is true for Tuc V, {  Phox II and Pic I,}
if kinematical studies
confirm its morphological identification as galaxies. Any further
discoveries towards the left of the $\tau_{DF}=5 Gyr$ curve would be
inconsistent with a particle dark matter hypothesis, specially if the systems detected,
as is the case for Dra II and Tuc V, were to show no morphological or kinematical indication
of being in the process of substantially reducing their sizes under dynamical friction.
{  If one assumes total dark matter halo sizes smaller by an order of magnitude, the coulomb
logarithm changes by $18 \%$, and the labels in the curves shown would change to
$5.9, 11.8$ and $17.7 Gyr$, the middle one still comparable or even shorter than the inferred
ages for these systems, often closer to $13 Gyr$ than to $10 Gyr$.

Table 1 gives a summary of the structural parameters and sources used for the figure, together with
four more systems which have been clearly identified as galaxies, but for which no published velocity
dispersion measurements exist. As these last are on average smaller than the four shown in the figure,
their inclusion would result in tighter constraints.
}

\section{Discussion}

{  The estimates for the dynamical friction evolution timescales presented here are not intended as
precise values, mostly due to the uncertainties in the density profile, distribution function
and overall sizes of the dark matter halos of ultra faint dSphs. An error of even an order of
magnitude in the halo size is quite possible, e.g. total $b_{max}$ values for the ultra faint
dSphs could be as low as $100 pc$, rather than the $1 kpc$ assumed here. This however, would only
change the value of $ln\Lambda$ by $log(10)$, from the 15 used above to 11.7. Also, the fiducial
ages of $10 Gyr$ taken here are in fact a lower limit, direct estimates of this parameter are
often slightly larger, ranging from $10 Gyr$ to $13 Gyr$ or even slightly more, also here with significant
uncertainties e.g. Drlica-Wagner et al. (2015). Thus, the most serious systematics probably
offset each other to some degree. In any case, in this first exploration of dynamical friction
in ultra faint dSphs it is not our intention to arrive at precise values for the decay timescales,
merely to note that the inferred values of the sizes and velocity dispersion values for the smallest
of the known systems, are already very close to the critical threshold for the problem. An interesting
result given that for all other known systems, dynamical friction of assumed dark matter halos
on individual stars are of hundreds and thousands of times the age of the universe.}

\begin{table}
\begin{flushleft}
  \caption{Parameters for the dSphs used.}
  \begin{tabular}{@{}lllllllll@{}}
  \hline
 \hline
   dSph &   Distance $(kpc) $ & $\sigma_{*} (km/s)$  & $R_{1/2} (pc)$  & References\\
 \hline
 & & & &  \\
 Dra II      &    $22$         & $2.9^{+2.1}_{-2.1}$    & $19^{+8.0}_{-6.0} $  & 5, 6 \\
 & & & &  \\
 Hor I       &    $79$         & $4.9^{+2.8}_{-0.9}$    & $30^{+4.4}_{-3.3} $  & 4    \\
 & & & &  \\
 Ret II      &    $30$         & $3.2^{+1.6}_{-0.5}$    & $32^{+1.9}_{-1.1} $  & 4    \\
 & & & &  \\
 Tri II      &    $36$         & $5.1^{+4.0}_{-1.4}$    & $34^{+9.0}_{-8.0} $  & 2    \\
 & & & &  \\
 Cet II      &    $30$         & Not known             & $17^{+7.0}_{-7.0} $  & 1    \\
 & & & &  \\
 Phox II     &    $83$         & Not known             & $26^{+6.2}_{-3.9} $  & 3    \\
 & & & &  \\
 Pic I       &    $114$        & Not known             & $29^{+9.1}_{-4.4} $  & 3    \\
 & & & &  \\
 Tuc V       &    $55$         & Not known             & $17^{+6.0}_{-6.0} $  & 1    \\
 \hline
\end{tabular} 

The table gives the structural parameters for the ultra faint dSphs appearing in the figure.
The last four rows show smaller systems presumably with shorter $\tau_{DF}$, which do not appear
in the figure as there are no reported velocity dispersion observations to date, although they have
been morphologically clearly identified as galaxies. The references used are: 1 Drlica-Wagner et al.
(2015), 2 Kirby et al. (2015b), 3 Koposov et al. (2015b), 4 Koposov et al. (2015a), 5 Laevens et al.
(2015) and 6 Martin et al. (2016).
\end{flushleft}
\end{table}

{ 
Another possible caveat to the results presented here lies in the probable reduction of dynamical
friction with respect to the Chandrasekhar formula within the constant density core of a dark matter
halo. As shown analytically in Hernandez \& Gilmore (1998), and confirmed numerically in various
studies, (e.g. Goerdt et al. 2006, Read et al. 2006, Inoue 2009, Cole et al. 2012) dynamical friction
on a perturber on a circular orbit will stall significantly within the constant density core region
of a dark matter halo. It would be extremely interesting to have detailed dynamical modelling of
the newly discovered dSphs, which might yield density profiles for their assumed dark matter halos
beyond the currently available bulk mass to light ratios. Also, the extent to which this effect might
play a part needs to be investigated through numerical simulations considering the particular
situation; a large number of randomly oriented stars on isotropic orbits.
}

We have shown that dark matter dominated pressure supported stellar systems will be subject to
dynamical friction on their individual stars, resulting in a substantial reduction of their typical radii
over a timescale of $\tau_{DF}=0.14 (\sigma_{*}/kms^{-1}) (r_{h}/pc)^{2} (ln\Lambda)^{-1}  Gyr$.
For typical inferred ages of dSphs, this dynamical friction shrivelling will become relevant 
for sizes of the order of the smallest recently detected ultra faint dSphs, with a theoretical 
lower stability limit of $\approx 19 pc$ on the smallest such system to be expected. Any future
detections of even slightly smaller ultra faint dSphs would be incompatible with a particle
dark matter hypothesis. This highlights the importance of obtaining more kinematical and structural
high accuracy determinations on these interesting systems {  to asses the relevance of the caveat noted
above}, and of looking for those statistically harder to find, the low size tail of the ultra faint dSph
distribution.

Under an alternative MONDian gravity scenario, no such consistency lower limits appear, as in 
the absence of dark matter, no dynamical friction ensues. Also, observed velocity dispersion
values for ultra faint dSphs agree with MONDian predictions of equilibrium velocities, 
for the measured stellar content of the systems in question.

\section*{acknowledgements}

The author acknowledges substantial constructive criticism from an anonymous referee as important
towards reaching a more complete and clear final version, and helpful input from Karina Voggel.
This work was supported in part by DGAPA-UNAM PAPIIT IN-100814 and CONACyT.

\end{document}